\documentclass[aps,twocolumn,groupedaddress]{revtex4}

\usepackage{graphics}

\begin{document}

\title{Background charges and quantum effects in quantum dots transport spectroscopy}

\author{M. Pierre}
\author{M. Hofheinz}
\altaffiliation{\emph{Present address:} University of California, Santa Barbara}
\author{X. Jehl}
\author{M. Sanquer}
\affiliation{CEA-INAC-SPSMS-LaTEQS, CEA-Grenoble}
\author{G. Molas}
\author{M. Vinet}
\author{S. Deleonibus}
\affiliation{DRT-LETI-D2NT-LNDE, CEA-Grenoble}

\date{\today}

\begin{abstract}
 We extend a simple model of a charge trap coupled to a single-electron box to energy ranges and parameters such that it gives new insights and predictions readily observable in many experimental systems. We show that a single background charge is enough to give lines of differential conductance in the stability diagram of the quantum dot, even within undistorted Coulomb diamonds. It also suppresses the current near degeneracy of the impurity charge, and yields negative differential lines far from this degeneracy. We compare this picture to two other accepted explanations for lines in diamonds, based respectively on the excitation spectrum of a quantum dot and on fluctuations of the density-of-states in the contacts. In order to discriminate between these models we emphasize the specific features related to environmental charge traps. Finally we show that our model accounts very well for all the anomalous features observed in silicon nanowire quantum dots.
\end{abstract}


\maketitle
\section{Introduction}
\label{intro}

\begin{figure}[t]
  \resizebox{1\columnwidth}{!}{%
  \includegraphics{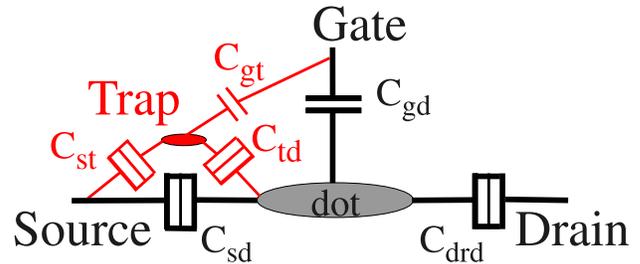} }
  \caption{Equivalent circuit for our model of a charge trap (red) coupled to a quantum dot (black). The trap occupation is 0 or 1 electron and it carries a very small current compared to the dot. The dot is treated within the orthodox model.}
  \label{schema}
\end{figure}

\begin{figure*}
\resizebox{\textwidth}{!}{%
\includegraphics{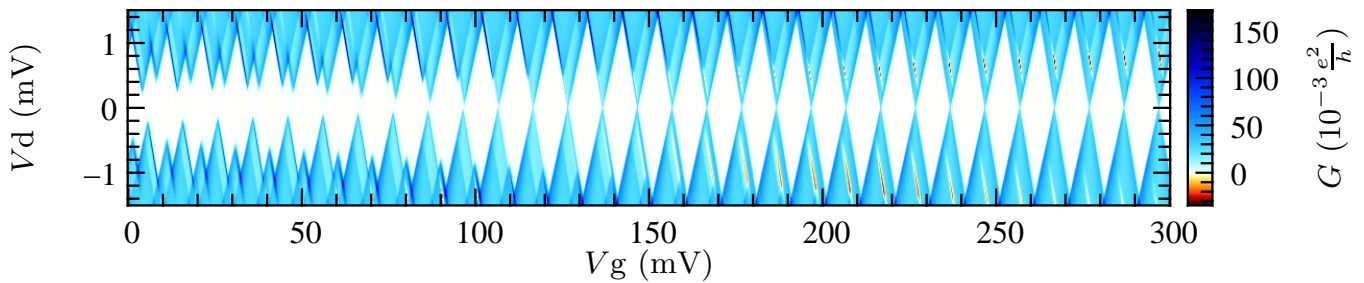} }
\caption{Simulated Coulomb diamonds (differential drain-source conductance versus drain and gate voltages) for our model depicted in Figure \ref{schema}. $C_\mathrm{gt}= 0.004\,\mathrm{aF} \ll C_\mathrm{gd}= 16\,\mathrm{aF}$, i.e.\ the  gate-trap capacitance is very small compared to the gate-dot capacitance. Near trap degeneracy ($V_g=0\,\mathrm{V}$) Coulomb diamonds are replicated because of the oscillating trap occupancy with $V_g$ (see Fig.~\ref{fig2Ter}). At larger $V_g$ the diamonds are shifted and undistorted; the trap is always occupied at zero bias but can still be empty at finite bias. This gives lines of differential conductance (see Fig.~\ref{fig2Bis}) slowly evolving with $V_g$.}
\label{fig2}
\end{figure*}

The diversity of nanostructures available for physicists to perform transport experiments at low temperature has grown considerably in the last decade. Following pionnering works on single-electron transistors (SETs) and quantum dots in the early 90's, it is now very common to observe and analyze Coulomb blockade in novel structures such as carbon based electronics or semiconducting nanowires.
We have developped in the past \cite{hofheinz06B} a silicon nanowire MOSFET which turns into a remarkably stable and simple SET below approximately 10\,K.  As it is widely accepted that background charges in the vicinity of the Coulomb island or the tunnel barriers are responsible for the large 1/f noise observed in many SETs, we developped a model based on a single charge trap capacitively coupled to the SET. This picture has been considered before in metallic single electron transistor, but without predictions for the Coulomb blockade spectroscopy \cite{Grupp}.
A peculiarity of our model is that the charge trap energy is not only sensitive to both source-drain and gate voltages, but also to the dot occupation number, because of the capacitive back-action of the dot upon the trap. We have shown that solving the rate equations for this model gives the same sawtooth pattern than observed experimentally \cite{hofheinz06}.
Sawtooth-like distortions of Coulomb diamonds are widely reported in the literature of quantum dots, for instance in carbon nanotubes SETs \cite{Cobden2}, graphene \cite{Ponomarenko}, molecules in gaps \cite{Heersche,Kubatkin} or epitaxial nanowires \cite{Thelander}.
As the data published on Coulomb blockade transport spectroscopy, either from us or other groups, have released more and more details, we have pushed our model further in order to see if the features observed by experimentalists and usually attributed to other effects can also arise from our model. The most common feature is certainly lines of differential conductance above the blocked region, parallel to the diamonds edges. These lines are often observed in undistorted diamonds, therefore one can conclude that our Background Charge (BC) model cannot apply in this case, as it predicts a clear distortion of the pattern.
In this work we show that with the right choice of capacitive couplings our BC model does predict these lines  in Coulomb spectra. It also gives negative differential conductance lines in a natural way without introducing adhoc hypothesis. We compare the BC model to the excitation spectrum (ES) model and the density of states (DOS) fluctuations model which are both usually invoked to explain such lines. We explain in details the origin of the most important features predicted by the BC model, the differences between the 3 models and how they can be distinguished. Finally we report data on silicon nanowire transistors where the traps are implanted arsenic dopants. We observe lines of differential conductance which are well reproduced by the BC model.

\section{The background charge model}
\label{sec:1}

In this paper we consider the general case represented in figure \ref{schema} of a quantum dot connected to a source, a drain and a control gate allowing to shift its electrostatic energy. In addition a charge trap occupied by 0 or 1 electron is located nearby. The current through the trap is negligible compared to the main current through the dot but the trap is weakly coupled by tunneling barrier to one contact and to the dot to allow its electron occupation to toggle between 0 and 1 electron. The energy of the trap is also fixed by the same gate voltage used to control the dot, but with a different capacitive coupling. This purely electrostatic model is a particular case of coupled quantum dots \cite{vanderwiel}. We chose arbitrarily to locate the trap on the source side. The source is grounded and a transport voltage is applied to the drain ($V_d$). We neglect any energy dependence of the tunnel transparencies. The dots are neither simply in series, as there is a finite transparency between the source and the dot, nor in parallel as there is no direct source-drain current via the trap.
The electrostatic model is fully characterized by the capacitive couplings (see Fig.~\ref{schema}):
$C_\mathrm{mn}$ 
is the capacitance between m and n where m,n=d (dot), t (trap), s (source), dr (drain) or g (gate).
$C_\mathrm{t}= C_\mathrm{td}+ C_\mathrm{st} +C_\mathrm{gt}$ is the total capacitance of the trap.
$C_\mathrm{d}= C_\mathrm{td}+ C_\mathrm{sd} +C_\mathrm{drd} +C_\mathrm{gd}$ is the total capacitance of the dot.
We consider sequential tunneling events driving the system into a finite number of states, defined by $(\mathrm{N}_\mathrm{dot},\mathrm{N}_\mathrm{trap})$, the occupation numbers of the dot and trap.
We deduce the drain-source current from the stationary occupation probability of the trap and the probability distribution of each charge state in the dot calculated with the rate equations. It is possible to null the dot-trap or source-trap current without changing the conclusions. This does not change the electrostatic scheme but changes the resulting differential conductance lines (not discussed here). The reference of electrostatic energies is arbitrary and does not affect our model which implies only energy differences between configurations.

Figure~\ref{fig2} shows the result for $C_\mathrm{gt}$, $C_\mathrm{gd}$, $C_\mathrm{td}$, $C_\mathrm{st}$, $C_\mathrm{d}$ $= 0.004$, $16$, $1$, $2$, $117$\,aF respectively i.e. for very small trap-gate capacitance. The current via the trap is 1\,pA, to be compared with the dot-source and dot-drain conductances of $0.1\,\frac{ e^{2}}{h}$ (0.4\,nA for $V_{d} = 10^{-4}\,\mathrm{V}$).
The level arm parameter for the trap, $ \alpha_{t}= \frac{ C_\mathrm{gt} }{ C_\mathrm{t}} \simeq 0.0013 $ is very small, about 10 times smaller than in ref. \cite{hofheinz06}. The number of successive distorted Coulomb diamonds increases as $\alpha_{t}$ decreases. A small $ \alpha_{t}$ means that the trap is much less sensitive to the gate voltage than to the dot occupation and drain voltage. As we will see later the limit $ \alpha_{t}\rightarrow 0$ is similar to the DOS model. A small $\alpha_{t}$ also implies that the differential conductance lines are parallel to the diamond edges. We have checked that larger $\alpha_{t}$ gives different slopes.
We set to 0 volt the degeneracy point where the trap mean occupation number is $\frac{1}{2}$. Near this point we observe distorted Coulomb diamonds: two replicas appear, corresponding to the two charge states of the trap. The shift in gate voltage for the replicated diamond is given by $\Delta
  V_\mathrm{g} = \frac{e C_\mathrm{td}}{C_\mathrm{gd}C_\mathrm{t}}$.
 The largest distortion $\Delta
  V_\mathrm{g} \simeq \frac{e}{2 C_\mathrm{gd}}$ occurs
    when $C_\mathrm{st}= C_\mathrm{td}$ \cite{hofheinz06}.
This occurs if the trap has equal couplings to source and dot. If the trap is more coupled either to the source or to the dot the apparent shift is reduced.

\begin{figure}
\resizebox{\columnwidth}{!}{\includegraphics{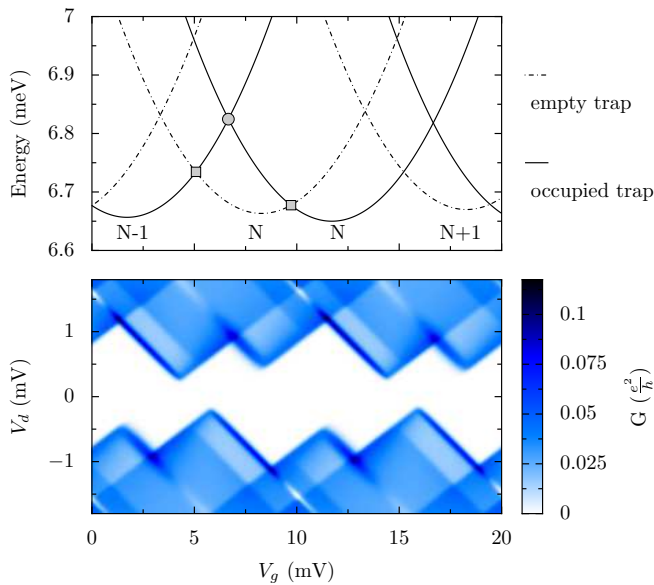}}
\caption{Same simulation than in figure~\ref{fig2}, focusing on the degeneracy region of the trap ($V_{g} \simeq 0\,\mathrm{V}$). Top panel: energy of lowest states at $V_d = 0\,\mathrm{V}$. Broken and solid lines are respectively for empty and occupied trap. Bottom panel: corresponding stability diagram. When increasing the gate voltage  the (N-1,1)$\rightarrow$(N,0) transition (left square) is reached before the (N-1,1)$\rightarrow$(N,1) transition (circle). The system then follows adiabatically the lowest energy state and the next transition is (N,0)$\rightarrow$(N,1) (right square). The (N-1,1)$\rightarrow$(N,1) transition is then replaced by  (N-1,1)$\rightarrow$(N,0)$\rightarrow$(N,1). The latter does not give a drain current.}
\label{fig2Ter}
\end{figure}

\begin{figure}
\resizebox{\columnwidth}{!}{  \includegraphics{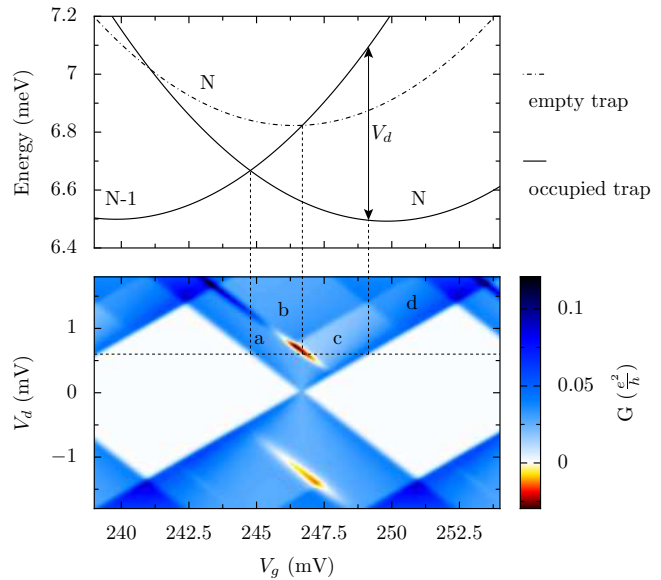}}
\caption{Same simulation than in figure~\ref{fig2}, focusing far from the degeneracy region of the trap ($V_{g} \gg 0\,\mathrm{V}$). Here the trap is always occupied at small bias voltage and the diamond is not distorted. Top panel: energy of lowest states at $V_d = 0.6\,\mathrm{mV}$.
Bottom panel: corresponding stability diagram.
Region \textit{a} : transitions (N-1,1)$\leftrightarrow$(N,1) can occur. Both states are sequentially obtained, hence a current flows through the dot.
Region \textit{c}: in addition, (N-1,1)$\rightarrow$(N,0) is also possible. When this occurs, no current passes through the dot anymore until the slow rate (N,0)$\rightarrow$(N,1) transition happen. The 0 state of the trap is a "dark state" lowering current through the dot. As a result a negative differential conductance line separates regions \textit{a} and \textit{c}.
Region \textit{b}: like in region \textit{c} the trap can exchange its electron with the dot, in addition the (N-1,0)$\leftrightarrow$(N,0) transitions can happen. Therefore, whatever the trap state current through the dot can circulate and there is more  current through the main dot than in \textit{a}, hence a positive differential conductance line separates regions \textit{a} and \textit{b}.
Region \textit{d}: the dot can even be filled with N+1 electrons when the trap is empty.}
\label{fig2Bis}
\end{figure}

A very important point is that the drain current vanishes at small bias near the degeneracy point, when the diamonds are replicated. Figure \ref{fig2Ter} illustrates the physical origin of the effect: Starting from $V_g=0\,\mathrm{V}$  at small bias ($V_d \simeq 0\,\mathrm{V}$) when the gate voltage increases an electron is transferred from the trap onto the main dot. This corresponds to the (N-1,1)$\rightarrow$(N,0) transition (left square in figure \ref{fig2Ter}), which occurs before the (N-1,1)$\rightarrow$(N,1) transition (circle on figure \ref{fig2Ter}) because of the repulsion by the charged trap. This charge exchange between the trap and the dot does not give any drain-source current.
More complex sequences with the same initial and final states and an intermediate state could in principle give a finite current. 
First, co-tunneling events are not taken into account. One could also expect an extra electron to tunnel first from the source onto the dot, then the trap releasing its electron to the source. Alternatively the latter could happen first, then an electron could tunnel from source to dot. However both sequences involve intermediate states too high in energy ((N,1) and (N-1,0) respectively), and therefore are forbidden. 
For the same reason, once in the (N,0) state it is impossible for an electron to exit the dot into the drain, because the (N-1,0) state is too high in energy at this gate voltage.
Increasing the gate voltage further, the system reaches the (N,0)$\leftrightarrow$(N,1) degeneracy (right square in Figure \ref{fig2Ter}). At this point the source can release an electron to the trap.
This second transition at constant number of electrons in the dot does not yield any source-drain current. In summary the (N-1,1)$\rightarrow$(N,1) transition has been replaced by the (N-1,1)$\rightarrow$(N,0)$\rightarrow$(N,1) sequence where the trap occupation oscillates in gate voltage by successive transfers from the source and into the dot.
In this gate voltage range the only way to recover a drain current is by applying a sufficient bias to allow N-1 or N electrons on the dot. Although a comparable situation has been described qualitatively in Ref. \cite{Tans}, our calculations do not predict the 'kinks' they observe, but replicas instead.

Far from the trap degeneracy ($V_g \gg 0\,\mathrm{V}$), our model recovers undistorted Coulomb diamonds, as shown in Figure~\ref{fig2}. Drain current is restored at low bias as (N-1,1)$\rightarrow$(N,1) is now the lowest energy transition.
More interestingly, our model predicts lines of differential conductance in the non blockaded regions, both negative and positive, as readily visible in Figure~\ref{fig2} and shown in more detail in Figure~\ref{fig2Bis}.
At some bias, the mean occupation of the trap is allowed to vary. The system can therefore be in various charge states, implying different total conductances, hence maxima of differential conductance.
We now discuss in more detail the origin of a negative differential conductance line  between regions labeled \textit{a} and \textit{c} in Figure~\ref{fig2Bis}.
As mentionned before, in both of these regions (N-1,1)$\leftrightarrow$(N,1) transitions are allowed and responsible for a finite drain current.
In addition, the (N-1,1)$\rightarrow$(N,0) transition is possible only in region \textit{c}, allowing the trap to transfer its electron to the dot.
Whenever this event happens no current passes through the dot anymore until the trap is filled again by an electron from the source.
The latter event is much slower than tunneling between source, dot and drain, so the current is smaller in region \textit{c} than in \textit{a} and therefore the differential conductance is negative.
The (N,0) state (trap empty) is a "dark state" blocking the current through the dot.
We have checked that slowing the (N,0)$\rightarrow$(N,1) transition by decreasing the source-trap tunneling rate reduces further the current in region \textit{c}.

The main conclusion of this study over a large gate voltage range (Figure~\ref{fig2}) is that both the negative differential conductance lines far from degeneracy and the suppression of current at low bias near degeneracy have the same origin, namely the electrostatic interaction between the trap and the dot. This effect can be exploited to build Coulomb blockade rectifiers suggested in ref. \cite{Stopa02,Likharev01}.

\section{Comparison with the ES and DOS models}
\label{sec:2}

Usual explanations for lines parallel to the edges of Cou\-lomb diamonds are based on the ES or DOS models.
Transport excitation spectra have been investigated in much detail \cite{zumbuhl,cobden} and their modification due to the Zeeman effect \cite{Cobden2}, photons \cite{Oosterkamp} or phonons \cite{Fujisawa} are widely observed in the Coulomb blockade stability diagram of quantum dots.

\begin{figure}
\resizebox{\columnwidth}{!}{%
  \includegraphics{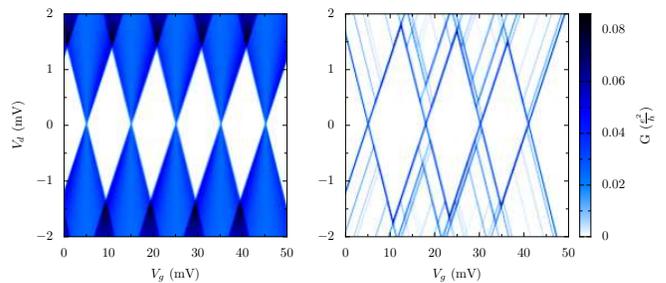}
}
\caption{Left panel: 
Simulation of the stability diagram (differential conductance versus gate and drain voltage) for a quantum dot, where the mean level spacing is negligible compared to the temperature T= 0.1K (orthodox model).
The charging energy is constant and equal to 1.38 meV ($C_\mathrm{d}$= 116 aF). Right panel: simulation of a dot with the same charging energy 
at the same temperature, but with 4 electrons occupying 4 non degenerate orbital states with spacings of 0.4, 0.3 and 0.1 meV respectively
(constant charging energy model).  Positive differential conductance lines appear outside the Coulomb diamonds. The one-particle excitation spectrum
shifts between successive diamonds as levels get
occupied, the excited state for N electrons becoming the ground state for N+1 electrons.}
\label{spectre_discret}     
\end{figure}

The ES Model is based on the existence of a resolved excitation spectrum because of the quantized kinetic energy of electrons confined in a small volume. It requires a temperature much lower than the mean energy level spacing $\Delta$.
We illustrate how lines appear in this model in figure \ref{spectre_discret}. The left panel is the orthodox model simulation of a quantum dot with $\Delta\ll k_BT$. The charging energy is constant and equal to 1.38\,meV ($C_d = 116\,\mathrm{aF}$). The right panel corresponds to a dot with the same charging energy, at the same temperature but with 4 electrons occupying 4 non degenerate orbital states with spacings of 0.4, 0.3 and 0.1\,meV respectively.
Positive differential conductance lines appear. In this constant charging energy model the one-particle excitation spectrum
shifts between successive diamonds as levels get occupied, the first excited state for N electrons becoming the ground state for N+1 electrons.
Very often scrambling of the one-particle spectrum after adding electrons is observed but recently a clear correlation between successive diamonds has been reported \cite{Ralph08}.
The absence of scrambling is interpreted as the absence of variation of the dot shape when adding electrons. That requires a steep enough confinement potential, for instance sharp etched edges.
In the ES model it is natural to expect only positive differential conductance lines because excited states (at higher energies than ground states by principle) are more tunnel coupled to the electrodes. If equal tunneling rates are supposed for the ground and excited states, as done in Figure \ref{spectre_discret}, then differential conductance is always positive. To observe negative differential conductance within the ES model the excited state must be less conducting. This could occur because of a specific fluctuation of the wavefunction envelop, a selection rule \cite{weinmann}, a blocking state \cite{datta} or Stark effect \cite{Dollfus06}.

The DOS model is appropriate if the very same pattern is observed in successive diamonds \cite{Falko97,Falko01}, in strong contrast with the ES model, because the energy in the electrodes does not depend on the gate voltage.
In this model the source-drain current is proportionnal to the local density of states $\nu_{S,D}(E)$ in the
source and drain. If for simplicity we suppose that the conductance is dominated by source-dot tunnel barrier, then $G(V)\propto {\frac{d\nu_{S}}{dE}}$.
In that case lines of differential conductance are expected to be parallel to the negative slope edge of the Coulomb diamond. This edge corresponds to the alignment of the last unoccupied level in the dot with the Fermi level in the source. In the opposite case of a dominating dot-drain barrier the lines will be parallel to the positive slope and $G(V)\propto{\frac{d\nu_{D}}{dE}}$.

In heavily doped semiconducting electrodes, just like in any diffusive reservoir, the local DOS fluctuates with the energy because of quantum interferences of elastically scattered quasi-particles diffusing coherently within a length scale related
to their lifetime at a particular energy \cite{Falko01}. The characteristic energy for these fluctuations is set by the inverse of the quasi-particle relaxation time, which decreases as the energy moves away from the Fermi energy.
The local DOS fluctuations increase when the diffusion coefficient $D=\frac{1}{d} v_{F}^{2}\tau$ decreases, with $d$ the dimensionality of the electrodes, $v_{F}$ the Fermi velocity and $\tau$ the elastic mean free time. Indeed a very large $D$ corresponds to a near perfect electronic reservoir without local fluctuations of the DOS.
The local DOS fluctuations are reinforced in small, confined geometries. If the trap considered in our model gets closer to the source, at some point electronic orbitals in the source and trap will overlap and a strong local fluctuation of the DOS appears due to hybridization. Then the energy of the trap depends less and less on the gate voltage. Therefore there is a continuous evolution from the DOS model (which is a purely quantum mechanical model) to our electrostatic model when increasing the trap-electrode distance.
The DOS model can explain other features which cannot be understood within the ES model \cite{Fasth}. First the DOS model naturaly explains negative differential conductance lines, as the sign of $\frac{d \nu_{S}}{dE}$ changes. In average one expects as many positive and negative differential conductance lines. Also the DOS model, can explain lines pointing to energies below the first available level in the dot. Such lines cannot arise from the ES model, as shown in the first  diamond in the right panel of Figure~\ref{spectre_discret}.

Our model differs from the ES and DOS models by the 3 following facts. 1) Lines are shifted progressively from one diamond to the next, in contrast to both ES and DOS models. 2) Lines coexist with sawtooth distortions (replicas) of diamonds at different gate voltages. 3) Negative differential conductance appears at finite bias voltage together with anomalously small current at low bias in distorted diamonds.
Like the DOS model it explains why lines can appear at energies extrapolated below the ground state of the artificial atom.
Finally our model, unlike both ES and DOS models, can explain lines not parallel to the diamond edges (if $\alpha_{t}$ is large) \cite{Cobden2,Zhong}.

\section{Silicon Quantum dot with arsenic donor as a trap}
\label{sec:3}

\begin{figure}
\resizebox{\columnwidth}{!}{%
  \includegraphics{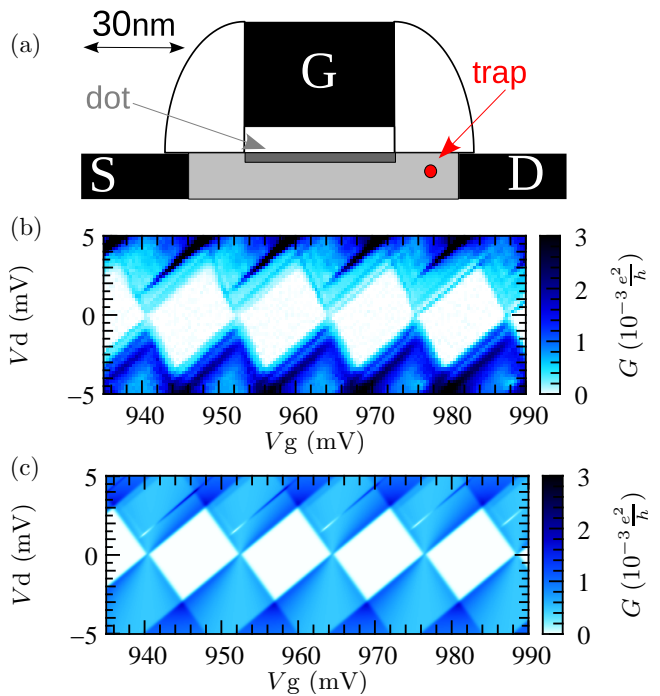}
}
\caption{(a) Schematic view of our MOS-SET. (b) Color plot of the drain differential conductance versus gate and drain voltages at T=350\,mK, which exhibits the same lines of conductance in successive diamonds with a small shift from diamond to diamond. These lines are explained with a purely electrostatic model involving a background charge. (c) Simulation with our model and parameters given in the text. We obtain a line at the same position than in the experimental data and which evolves slowly with $V_g$ because the trap is weakly coupled to the gate.}
\label{fig6}
\end{figure}

We originally developped our model for silicon nanowire transistors with implanted arsenic traps in the tunnel barriers \cite{hofheinz06}.
Here we report new data recorded in similar samples which clearly show the correlations in successive undistorted
 Coulomb diamonds predicted by our model.
Fig.~\ref{fig6}b shows a typical stability diagram with a pattern of differential conductance lines very weakly dependent on the number of carriers in the dot.
This weakly shifted pattern is a characteristic signature of our model for traps much more coupled to the source (or drain or dot) than to the gate.
The samples are described in ref. \cite{hofheinz06B,hofheinz06,thesemax} and schematically drawn in Fig.~\ref{fig6}a.
They are SOI-MOSFETs adapted in terms of doping to become controlled SETs at low temperature. A 20 to 80\,nm wide wire is
etched to form the channel. The source and drain parts of the wire are highly doped to form good metallic reservoirs (As,
$\simeq 10^{20}\,\mathrm{cm}^{-3}$). The central part of the wire is covered by a 40\,nm poly-Si gate electrode, isolated by
Si0$_\mathrm{2}$, and self-aligned silicon nitride spacers (35\,nm long) are deposited on both sides of the gate (rounded walls in Fig.~\ref{fig6}a). The part of the wire below the gate and spacers (light gray regions on Fig.~\ref{fig6}a) is only lightly doped (As,
$5\times 10^{17}\,\mathrm{cm}^{-3}$), so that at low temperature it forms an insulator. However directly below the gate electrode it
can be tuned into a metallic state by applying a positive gate voltage. That way a quantum dot is formed under the gate. The tunnel barriers are the low-doped parts of the wire adjacent to the dot. The arsenic dopants inside these tunnel barriers are weakly capacitively coupled to the gate and are the traps considered in our model. Well centered donors give replica for the diamonds \cite{hofheinz06}.
There is a gradient of Arsenic concentration (typically one order of magnitude for a 5 nm lateral distance in our process) at the border between the source-drain and channel. Many arsenic donors are located close, and are therefore well coupled to the source or drain. Such donor have a small lever arm parameter $ \alpha_{t}$ and give undistorted diamonds with lines of differential conductance. Donors which are strongly coupled to the dot (due to electrostatic bending of the impurity band below the gate edges) produce the same effect.

We performed our measurements at 350\,mK in a $^\mathrm{3}$He refrigerator and measured the differential conductance with a standard ac lock-in technique. At this temperature we do not expect to resolve the quantum levels in the dot. The mean energy level spacing $\Delta$ between quantum states is the largest for small dots at low gate voltages where only a 2D electron gas is formed at the
surface of the channel. In this limit we expect $\Delta_\mathrm{2D} \sim \frac{2 \pi \hbar^2}{d m^* A} \sim
150\,\mu\mathrm{eV}$, with $d=4$ the spin and valley degeneracy, $m^*=0.19\,m_e$ the 2D effective mass, and $A \simeq
4000\,\mathrm{nm}^2$ the total surface area of the gate/wire overlap, including the flanks of the wire. As the dot gets filled, the electron gas eventually fills up the whole volume of the wire below the gate and $\Delta$ falls below $20\,\mu\mathrm{eV}$. Quantum levels can only be resolved when $\Delta$ is larger than the width of the resulting lines of differential conductance. These lines have a full width at half maximum of approximately $3.5\,k_\mathrm{B}T \sim 100\,\mu\mathrm{eV}$ given by the Fermi
distribution in the leads. For the large gate voltages shown in figure~\ref{fig6} we are in the high density regime where $\Delta$ is too small to play a significant role. Therefore the sharp lines of differential conductance seen on figure ~\ref{fig6} cannot be explained within the ES model. The lines have a typical energy separation of 1\,meV, much larger than the calculated mean spacing of $20\,\mu\mathrm{eV}$.
They are also observed at larger energy than expected in the DOS model. In our samples the reservoirs are highly doped silicon wires in which the mean level spacing is very small, and local DOS fluctuations have a correlation in energy of the order of the inverse inelastic time $\simeq 1\,\mathrm{ns}$ (${\frac{h}{\tau_{in}}} \simeq  4\,\mu\mathrm{eV}$) at T=1K in heavily doped silicon \cite{heslinga}.

In summary, unlike the ES and DOS models, our model explains quantitatively the weakly shifting lines between successive diamonds measured in Figure~\ref{fig6}b, with a trap located on the drain side of the dot. Figure~\ref{fig6}c shows the result of the simulation with $C_\mathrm{gt}$, $C_\mathrm{gd}$, $C_\mathrm{td}$, $C_\mathrm{drt}$, $C_\mathrm{d} = 0.006$, $13.3$, $0.4$, $0.046$, $53.3$\,aF.

\section{Conclusions}
\label{conclusion}

We have extended a simple electrostatic model of a charge trap coupled to a dot to the case of very weak coupling to the gate. In this regime new features are predicted over a large gate voltage range. Near degeneracy of the trap the sawtooth pattern calculated in a previous work is recovered and the current suppression at low bias voltage is understood in more detail. We obtained new features far from this degeneracy, where Coulomb diamonds are not distorted. Lines of differential conductance appear in the diamonds, very similar to the ones usually attributed excited states, although our model does not involve a discrete spectrum for the dot.
These differential conductance lines can be positive or negative, and parallel to either edge of the diamond. They also evolve very weakly with gate voltage, an original feature not predicted by other models.
Our model easily accounts for negative differential conductance lines, a feature usually attributed to density-of-state fluctuations in the contacts. This model and ours are converging when considering a trap located very close to the electrodes, as hybridization occurs and coupling to the gate goes to zero.
Although the most basic signature of our model is the sawtooth pattern and associated current suppression, we emphasize that its experimental observation is not required to validate our charge trap scheme. Indeed lines can very well be observed in undistorted diamonds while the degeneracy region remains out of the energy range which can be probed experimentally.

Even though our model involves only a single trap occupied with zero or one electron and a dot treated in the orthodox model, it already gives a complicated pattern of lines and features in the stability diagram. 
It provides a quantitative but simple way to simulate the Coulomb blockade spectroscopy of quantum dots, and shows the great impact of a single charge on this spectrum.
Further extensions like several traps, resolved mean one-particle level spacing in the dot, non-negligible current through the trap, double occupation of the trap, Zeeman effect on the trap energy can be implemented in the near future.
As more and more nanostructures designed for transport experiment exhibit Coulomb blockade, our model could account for many features observed experimentally as the presence of charge traps is very realistic.


\begin{thebibliography}{}


\bibitem{hofheinz06B} M. Hofheinz, X. Jehl, M. Sanquer, G. Molas, M. Vinet and S. Deleonibus, Appl. Phys. Lett. \textbf{89}, (2006) 143504.
\bibitem{Grupp} D. E. Grupp, T. Zhang, G. J. Dolan and N. S. Wingreen, Phys. Rev. Lett. \textbf{87}, (2001) 186805.
\bibitem{hofheinz06} M. Hofheinz, X. Jehl, M. Sanquer, M. Vinet, B. Previtali, D. Fraboulet, D. Mariolle and S. Deleonibus, Eur. Phys. J.  \textbf{B 54}, (2006) 299.
\bibitem{Cobden2} D. H. Cobden and J. Nyg\aa{}rd, Phys. Rev. Lett. \textbf{89}, (2002) 046803.
\bibitem{Ponomarenko} L. A. Ponomarenko, F. Schedin, M. I. Katsnelson, R. Yang, E. W. Hill, K. S. Novoselov and A. K. Geim, Science, \textbf{320}, (2008) 356.
\bibitem{Heersche} H. B. Heersche, Z. de Groot, J. A. Folk, H. S. J. van der Zant, C. Romeike, M. R. Wegewijs, L. Zobbi, D. Barreca, E. Tondello and A. Cornia, Phys. Rev. Lett. \textbf{96},  (2006) 206801.
\bibitem{Kubatkin}S. Kubatkin, A. Danilov, M. Hjort, J. Cornil, J.-L. Br\'edas, N. Stuhr-Hansen, P. Hedeg\aa{}rd, T. Bj\o{}rnholm, Nature \textbf{425}, (2003) 698.
\bibitem{Thelander} C. Thelander, T. Martensson, M. T. Bjork, B. J. Ohlsson, M. W. Larsson, L. R. Wallenberg, and L. Samuelson, Appl. Phys. Lett. \textbf{83}, (2003) 2052 .
\bibitem{vanderwiel} W. G. van der Wiel, S. de Franceschi, J. M. Elzerman, T. Fujisawa, S. Tarucha and L. P. Kouwenhoven, Rev. Mod. Phys. \textbf{75}, (2003) 1.
\bibitem{Tans} S. J. Tans, M. H. Devoret, R. J. A. Groeneveld and C. Dekker, Nature \textbf{394}, (1998) 761.
\bibitem{Stopa02} M. Stopa, Phys. Rev. Lett. \textbf{88}, (2002) 146802. 
\bibitem{Likharev01} S. Folling, O. Turel, and K. K. Likharev,  "Single-Electron Latching Switches as Nanoscale Synapses", in: Proc. IJCNN01 Neural Networks,(2001) 216.
\bibitem{zumbuhl} D. M. Zumbuhl, C. M. Marcus,  M. P. Hanson and A. C. Gossard, Phys. Rev. Lett. \textbf{93}, (2004) 256801.
\bibitem{cobden} D. H. Cobden, M. Bockrath, P. L. McEuen, A. G. Rinzler and R. E. Smalley, Phys. Rev. Lett. \textbf{81},  (1998) 681.
\bibitem{Oosterkamp} T. H. Oosterkamp, L. P. Kouwenhoven, A. E. A. Koolen, N. C. van der Vaart and C. J. P. M. Harmans, 
Phys. Rev. Lett. \textbf{78}, (1997) 1536.
\bibitem{Fujisawa} T. Fujisawa, T. H. Oosterkamp, W. G. van der Wiel, B. W. Broer, R. Aguado, S. Tarucha and L. P. Kouwenhoven, Science \textbf{282}, (1998) 932.
\bibitem{Ralph08}F. Kuemmeth, K. I. Bolotin, S.-F. Shi, and D. C. Ralph Nano Lett. \textbf{8} (12), (2008)  4506.
\bibitem{weinmann} D. Weinmann, W. Haeusler, B. Kramer, Phys. Rev. Lett. \textbf{74},  (1995) 984.
\bibitem{datta} B. Muralidharan and S.  Datta, Phys. Rev. \textbf{B 76}, (2007) 035432.
\bibitem{Dollfus06} J. S\'{e}e, P. Dollfus, and S.  Galdin IEEE Trans. on Elec. Dev.,  \textbf{53},  (2006) 1268.
\bibitem{Falko97} T. Schmidt, R. J. Haug, V. I. Fal'ko, K. V. Klitzing, A. F\"{o}rster and H. L\"{u}th, Phys. Rev. Lett. \textbf{78}, (1997) 1540.
\bibitem{Falko01} J. Koenemann, P. Koenig, T. Schmidt, E. McCann, Vladimir I. Falko, and R. J. Haug Phys. Rev.  {\bf B 64}  (2001) 155314.
\bibitem{Fasth} C. Fasth, A. Fuhrer, L. Samuelson, V. N. Golovach and D. Loss, Phys. Rev. Lett. \textbf{98}, (2007) 266801.
\bibitem{Zhong} Z. Zhong, Y. Fang, W. Lu and C. M. Lieber, Nano Lett. \textbf{5}, (2005) 1143.
\bibitem{thesemax} PhD thesis, M. Hofheinz, universit\'{e} J. Fourier 2006, Grenoble, unpublished (http://tel.archives-ouvertes.fr/tel-00131052/en/)
\bibitem{heslinga} D. R. Heslinga and  T. M. Klapwijk, Solid State Comm. \textbf{84}, (1992) 739.


\end{thebibliography}
\end{document}